# To Ban or Not to Ban: Uses and Gratifications of Mobile Phones among Township High School Learners

Khaya Kunene & Pitso Tsibolane
University of Cape Town, Cape Town, South Africa
knnkha003@myuct.ac.za
pitso.tsibolane@uct.ac.za

## *Abstract*

The proliferation of mobile phone usage among learners from diverse socio-economic backgrounds has prompted school authorities to contemplate banning these devices within educational institutions. This research seeks to explore the motivations and usage patterns of high school learners in response to the proposed ban. Employing a mixed-methods approach, we conducted surveys and interviews with 262 students from three township schools in the Western Cape province of South Africa. Grounded in the Uses and Gratification Theory (UGT), our study examined four key categories: reasons for mobile phone use, usage patterns, purchasing influences, and behavioral factors. Our findings reveal a predominant opposition among students to the ban, despite a significant number opting to leave their phones at home due to concerns about theft and robbery in their neighborhoods. Financial constraints, specifically the inability to afford data bundles and airtime, also contribute to this behavior. Notably, 40% of the participants reported using their phones for more than five hours daily, a duration classified as overuse in existing literature. The primary motivations for mobile phone use among these learners include socializing, internet browsing for non-educational purposes, and using the device for entertainment and recreation. This study highlights critical insights into the nuanced relationship between high school learners and mobile phone usage, offering valuable perspectives for policymakers and educators considering the implications of a mobile phone ban in schools.

# 1. Introduction

The use of mobile phones by learners has become a highly contentious topic between school authorities, parents and students worldwide (Gao et al., 2014, Van Der Ross & Tsibolane, 2017). The explosion of mobile texting has been cited as a source of distraction in classrooms for learners who tend to constantly text and also sext each other (Porter et al., 2015; Lenhart et al., 2010). The potential for cyberbullying also increases with the use of mobile phones among teenagers (Holfeld, 2012). The increased screen time, cheating by learners, accessing inappropriate content and the related low performance by the learners at school are also some of the reasons why mobile phones are viewed negatively particularly in school settings (Hong et al., 2012; O'Bannon & Thomas, 2014).

Despite the negative aspects of mobile phone usage by learners, other researchers have found positive benefits. Traxler (2009) argues that mobile phones can be beneficial as they enable learners to engage various learning opportunities ubiquitously. Mobile phones are considered to afford functionalities such as cameras and recorders that can assist learners to create new content (Hartnell-Young & Vetere, 2008). Mobile phones also enable students to learn collaboratively (Corbeil & Valdes-Corbeil, 2007) while enabling teachers to provide a targeted and differentiated learning environment (Kukulska-Hulme, 2007).

The purpose of this study is to assess how and why high school learners in disadvantaged areas of South Africa use mobile phones. The research questions are as follows:

- What are the reasons learners use mobile phones?
- How do learners use mobile phones?

To answer these research questions, the study adopts the Uses and Gratification theoretical model by Balakrishnan & Raj (2012) to identify 1) the reasons why learners use mobile phones, 2) patterns of mobile phone usage, 3) purchasing factors as well as 4) the behaviour related issues that are associated with using mobile phones. A survey of 262 high school students from three township schools in the Western Cape Province of South Africa

The rest of the paper is divided into four sections. Section 2 will explain the constructs related to UGT as adopted in this study from literature. This will be followed by the theoretical model discussion in Section 3. Section 4 focusses on the analysis of the data. Section 5 summarises the findings. Section 6 concludes the paper and suggests future work.

# 2. Literature Review: A discussion of the related constructs

This literature review explores the mobile phone policy in high schools. This is followed by a discussion of the identified uses and gratifications of mobile phones by learners in high schools.

## 2.1 Mobile Phone Policy

Mobile phone ban school policies are common in South Africa with about 85% of learners reporting that they are not allowed to bring phones to school despite 41% of them disclosing to have brought a phone to school despite the ban (Porter et al.,2015). Another study found that about 78% of the learners in the study owned a mobile phone and about 57% of them regularly carry them to schools despite 71% of the same learners agreeing that phones should be banned due to their distractive influence (Maphalala & Nzama, 2014). Schools in the United Kingdom and the United States are beginning to revise the banning of mobile phones in schools due to the belief that in the 21st century, the use of technology can be managed thoughtfully to improve a learner and teacher productivity (Gao et al., 2017, 2014; Murphy, 2015).

The use of mobile phones by teachers in class also raises questions, because it is hard to differentiate between professional and non-professional use during a lesson (Porter et al., 2015). When a learner is seen using a mobile phone, the phone will be confiscated away either by a parent or the school and given back at the end of the year (Lenhart et al., 2010). There generally does not seem to be a uniform and consistent approach to policy planning with regards to the use of mobile phones and access to internet websites in South African schools (Conradie & Roodt, 2013; Ives, 2012).

## 2.2 Reason for mobile phone use among learners

Mobile phone technology has transformed how the young generation interact and engage with each other via social networks (Dube, 2014). The youth use mobile phones to communicate with family or fellow friends through phone calls and via social media at any given time without boundaries (Nawaz & Ahmad, 2012). The new generation of young people are always posting and processing multimedia content, such as text, audio, pictures and video's via Facebook, WhatsApp, Twitter and YouTube (Lederer, 2012; O'Hagan, 2015).

Even though parents do not generally condone the use of mobile phone during school hours, one of the reasons they acquire mobile phones for their children is security i.e. the ability to get hold of the children during emergency situations at all times (Balakrishnan & Raj, 2012; Sundari, 2015). Owning a mobile phone has become a status among the youth (North et al., 2014). Accessibility to internet and social networking has been one of the reasons learners own mobile phones (O'Hagan, 2015; Rambe & Bere, 2013; Takalani, 2008).

Despite having access to computer labs with internet connection, majority of the learners often bring and use mobile phones while at school in order to sustained constant communications (Ferriter & Garry, 2010). The majority of mobile phones used by learners are equipped with a number of features, such as a camera, music player and faster internet connection modem (Jones et al., 2010). Mobile phones help learners to maintain privacy within their circle of friends to the exclusion of parents and other unwelcomed authorities who are uninformed about the activities learners often plan to engage in (Balakrishnan & Raj, 2012). However, despite a growing need for smart phones with advanced phone features, factors such as crime, the maintenance of the phone and getting access to unlimited data remains a source of frustration for phone ownership for learners (Walls et al., 2014).

## 2.3 Mobile phone use patterns among learners

Mobile phone usage pattern refers to the amount of calls and text messages created or received within a given period of time (Balakrishnan & Raj, 2012). North et al (2014) observes that the pattern of mobile usage differs between females and male students, for example, females tend to use chat services on a greater scale compared to their male counterparts (North et al., 2014). An average learner makes between one to five phone calls a day between family members and friends (Balakrishnan & Raj, 2012). Due to high costs of making phone calls particularly in South Africa, the majority of young people prefer communication over social media by texting and sending voice notes (Dube, 2014).

## 2.4 Behaviour related factors

Learners with smartphones have been found to spend more time checking and using their devices compared to learners with feature phones (Balakrishnan & Raj, 2012; Nawaz & Ahmad, 2012; Sundari, 2015). Despite the strict rules that phones should not be used in class, learners still find time and opportunity to use or check their phones during lessons when the lecture is not insight (Maphalala & Nzama, 2014). Learners also tend to get attached to their mobile phones and often feel stressed when mobiles are out of reach (Balakrishnan & Raj, 2012). The majority of learners have been found to be addicted to their mobile devices (North et al., 2014).

## 2.5 Reasons for purchasing mobile phones

There are various reasons why learners purchase or choose a particular mobile device. Factors such as the type of a phone brand, functionality, value, price and phone features such as touch screen and storage capacity are key to the decision (Balakrishnan & Raj, 2012). Majority of learners from disadvantaged communities often own mobile phones or device instead of a computer or laptop due to cheaper maintenance cost (North et al., 2014).

## 2.6 The negative effects of the over-use of mobile phones

Excessive use of mobile phone and browsing of internet can negatively affect a learner's health particularly as it relates to sleeping patterns which can lead to various health issues (Furze, 2014). Young people have become highly addicted to their mobile phones as they spend significant amounts of time per day chatting via social media (Hong et al., 2012; Porter et al., 2015). There is a growing concern that learners are becoming dependent on the use of web content (which could be unreliable due to the proliferation of unreliable and unverifiable internet sources) on their phones and no longer use libraries or traditional book material as primary sources of information (Oblinger et al., 2005). Even though this generation of learners believe they can multitask, they are often encouraged and required to listen carefully to the teacher as unnecessary disruption are not allowed in class (Ives, 2012; Takalani, 2008). Social media texting styles such as the use of slang also known as informal writing is often mistakenly used by the learners when writing assessments and in formal academic activities (O'Bannon & Thomas, 2015).

# 3. Research Model

Communication theories have evolved over the years. The Uses and Gratifications Theory (UGT) has primarily been used in the media environment such as in radio, press and television as a lens to explain various reasons why people tend to use specific new technologies (Ruggiero, 2000; Sanz-blas, 2014). The theory has been used to help examine the reasons people use technology or media to help meet and gratify individual social needs (Balakrishnan & Raj, 2012). Ruggiero (2000) states that the UGT was first used in the 1940s by the researchers as a tool to help explain why people used different forms of media, such as reading the newspaper, watching the television or listening to the radio. In the past, the theory has been adopted by researchers with a specific focus to a group of people who are active media users (Jahn & Kunz, 2012). The theory enables researchers to examine the motives behind the desire by consumers to use media and mobile phones to satisfy the needs that individuals actively seek (Sanz-blas et al., 2014). The theory has been criticised for failing to differentiate between gratifications sought and gratifications obtained (Sanz-blas et al., 2014). Balakrishnan and Raj (2012) further developed the UGT to assess the use of mobile phone by the youth in Malaysia. The model contains four main categories, namely: reasons to use mobile phones, pattern of mobile phone usage and behaviour related issues. The model has been used by North et al. (2014) and Eybers et al. (2015) to assess the why and how groups of young people use their mobile phones.

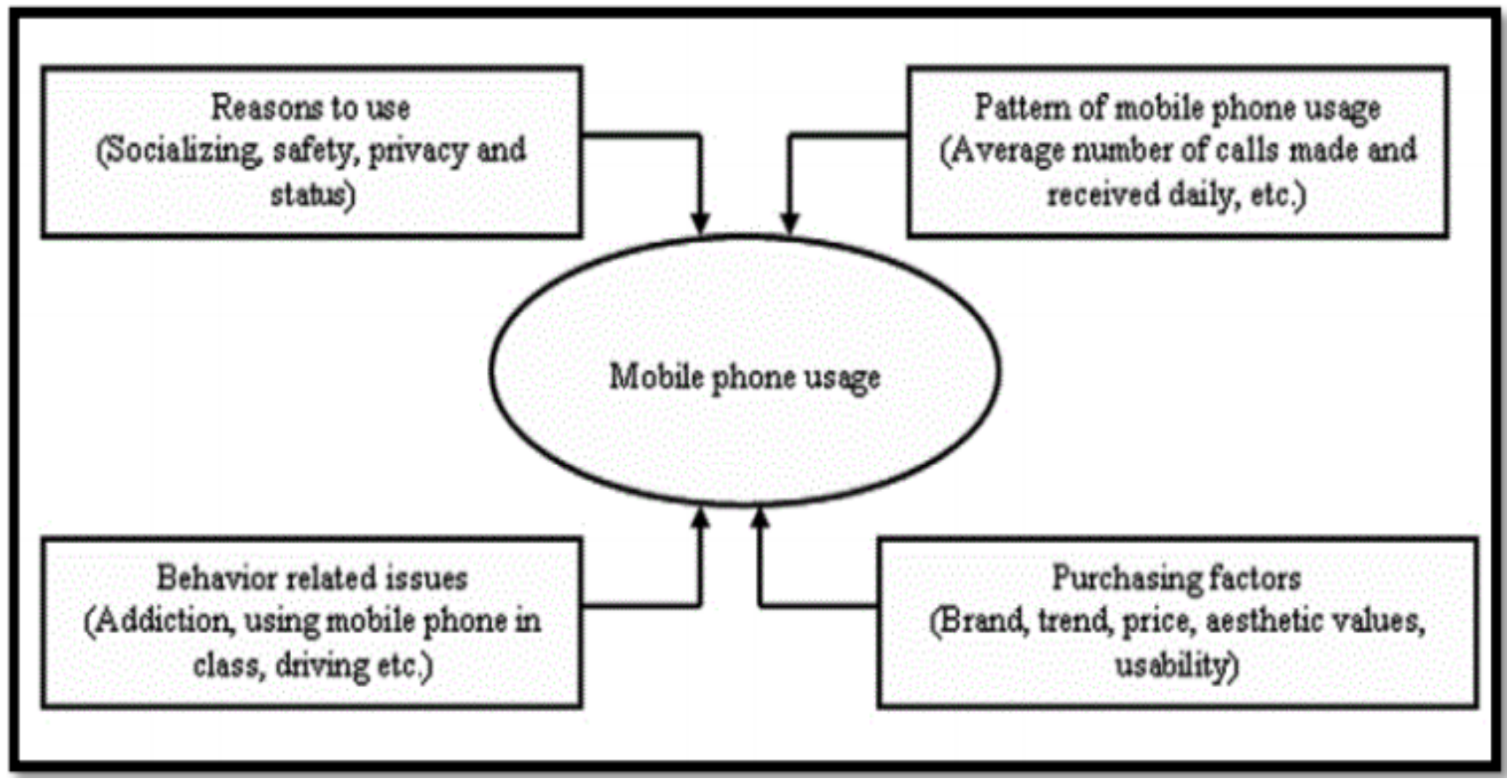

**Figure 1: Uses and Gratification Framework (Balakrishnan & Raj, 2012)**

# 4. Methodology

The study was conducted in year 2016 in the Western Cape, the data was collected through a paper-based questionnaire and interviews with Grade 12 learners (final high school year). The paper-based questionnaire was used due to the concern that learners in disadvantaged schools may not have access to online-based tools. A total of 35 questions were used in the questionnaire to assess how learners used mobile phones. The first section was about general information, second section was about the learner's exposure on mobile phones, third section asked about the type of mobile phones learners used and the last section used a Likert-scale starting from strongly disagree to strongly agree, then at the end the survey has open ended questions to allow the participants to freely express their views.

The research was conducted at three different high schools located in Cape Town. The survey received a total response of 262 learners across the three schools. A consent form was attached along with the survey for participants to read and sign. Among the chosen schools, 90 learners were from School A, 106 from School B and 66 from School C.

# 5. Data Analysis and Results

The study was conducted from three secondary schools within the western Cape, a total of 262 learners took part in the study (112 male & 144 female). A breakdown of the number of learners from each school who took part in the study are shown on the table 1 below.

| | High School A | | High School B | | High School C | | TOTAL |
|---|---|---|---|---|---|---|---|
| | *N* | *%* | *n* | *%* | *N* | *%* | |

| **GENDER** | | | | | | | |
|---|---|---|---|---|---|---|---|
| Male | **36** | *40* | **46** | *43,40* | **30** | *45,45* | **112** |
| Female | **53** | *58,89* | **57** | *53,77* | **34** | *51,52* | **144** |
| Unspecified | **1** | *1,11* | **3** | *2,83* | **2** | *3,03* | **6** |
| **AGE** | | | | | | | |
| 16-18 | **56** | *62,22* | **41** | *38,68* | **29** | *43,94* | **126** |
| 19-21 | **25** | *27,78* | **37** | *34,91* | **26** | *39,39* | **88** |
| Prefer not to say | **9** | *10* | **28** | *26,42* | **11** | *16,67* | **48** |
| **TOTAL** | **90** | ***100%*** | **106** | *100%* | **66** | *100%* | **262** |

**Table 1. Data analysis - Demographics**

Of the 262 learners surveyed, females counted for 53.83%, while the male counterparts counted for 44.27% and 1.91% of the participants preferred not to identify their gender. The youngest learner was 16 years of age and the oldest was 23 years. About 34% of the learners currently in Grade 12 are between the ages of 19 and 21 years. Only 7% of the learners did not own a mobile phone of their own but still had occasional access to a phone through a friend or family member. About 78% of those who owned a mobile phone possessed one device and about 6% possessed three phones. All the surveyed students owning a phone had a smartphone with Android being the most common (53%) mobile platform followed by Windows mobile phone (23%), Blackberry (11%) and the rest being iOS and other unspecified mobile platforms (13%).

## 5.1 Reason for mobile phone use

Other studies have found out that the most common reason to use a mobile phones among young people is for social status and for safety reasons such as calling parents for emotional advice (North et al., 2014). Our study has found out that the main reasons learners at the disadvantaged high schools surveyed use their mobile phones is for socialising (communicating and chatting with friends and family via WhatsApp, Facebook and voice), browsing the internet for general (non-school related) information as well as using the phone as an entertainment and recreational device (music and taking pictures). Learners also agreed to using their phones for school activities (searching for answers). Despite social media dominance when it comes to usage purpose amongst learners, majority of the learners often use their mobile phones as an everyday gadget that help learners operate and manage daily routines effectively, for example setting up an alarm, creating a to-do list, using the phone camera and also listening to music (Porter et al., 2015).

## 5.2 Mobile Phone Usage Patterns

Studies have revealed that the majority of learners use a mobile phones at least at an average of five ours a days (Balakrishnan & Raj, 2012). An average student spends nearly 2 to 6 hour a day on social media, (Porter et al., 2015; Unicef, 2012). In this study about 40% learners use their phones for more than five hours a day either viewing or checking missed called or message. Approximately 31% of learners use their mobile phones between the hours of three to four a day.

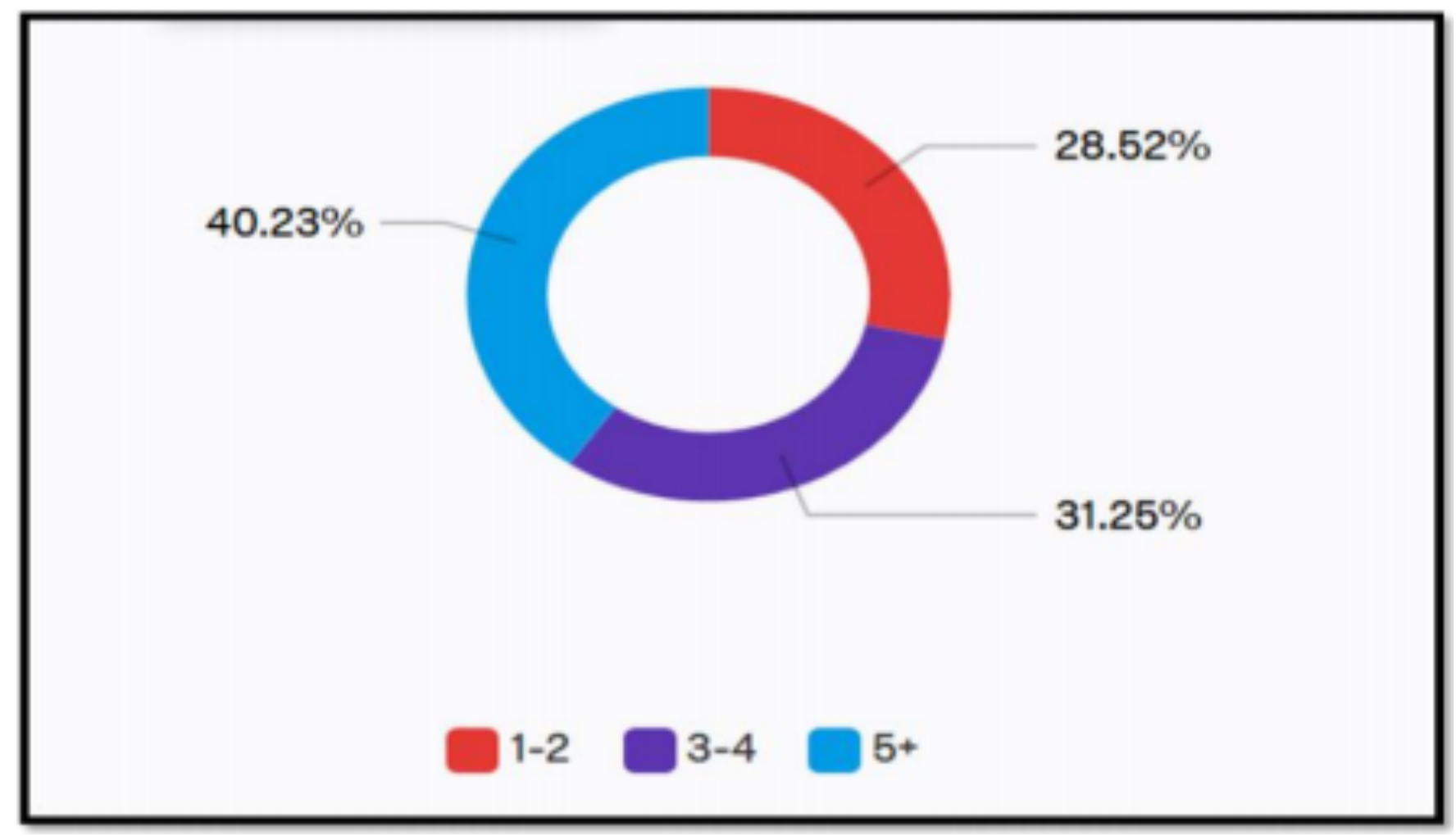

**Figure 2: How many hours do you spend using a mobile?**

## 5.3 Purchasing Factors

North et al. (2014) concludes that when purchasing a mobile phone, learners prioritise factors such as price, usability, brand and styling (look and colour). Our study revealed that the most significant factors for purchasing a mobile phone is affordability as well as phone features, particularly camera quality and storage capacity for multimedia content such as music and video.

## 5.4 Behavioural Issues

Previous studies have shown that when it comes to behavioural issue, many learners tend to be very attached to their mobile phones(Balakrishnan & Raj, 2012), while North et al. (2014) observed that learners often get stressed when their mobile phones are out of reach and access. In all the three schools surveyed in this study, a mobile phone ban policy was in place. However, it emerged in the interviews that the students are actually only leaving their phones at homes due to the following reasons:

- The fear of losing the phone due to theft at school
- Fear of being robbed by criminals on the way to school
- Lack of money to buy internet data bundles

This finding is in agreement with the findings of Maphalala and Nzama (2014) that on of the major reasons for learners to leave mobile phones at home was due to the concern of safety primarily and school ban secondarily.

### 5.5 Learner attitudes and feeling towards mobile phone use at school

Educational mobile phone studies have shown that even though mobile phones are helpful towards learning, students sometimes complain that mobile phone can be distracting and invasive during class (Sundari, 2015). Learners in our study also stated that mobile phones can be a distraction to learning as other learners could receive calls during class or can be seen chatting on the phone under the desk while pretending to pay attention to learning. When learners were asked "Should mobile phone be allowed at school", 56% of the respondents answered YES and 42% answered NO. Others (2%) advocated for a conditional position i.e. allowing phone use under strict rules, mainly for school-related purposes. Despite the acknowledgment by learners that mobile phones are a distraction within the school educational setting, many learners highlighted that their mobile phones have also provided them with other benefits such as access to online resources, dictionary and low cost communication with the peer learners (Maphalala & Nzama, 2014; Porter et al., 2015).

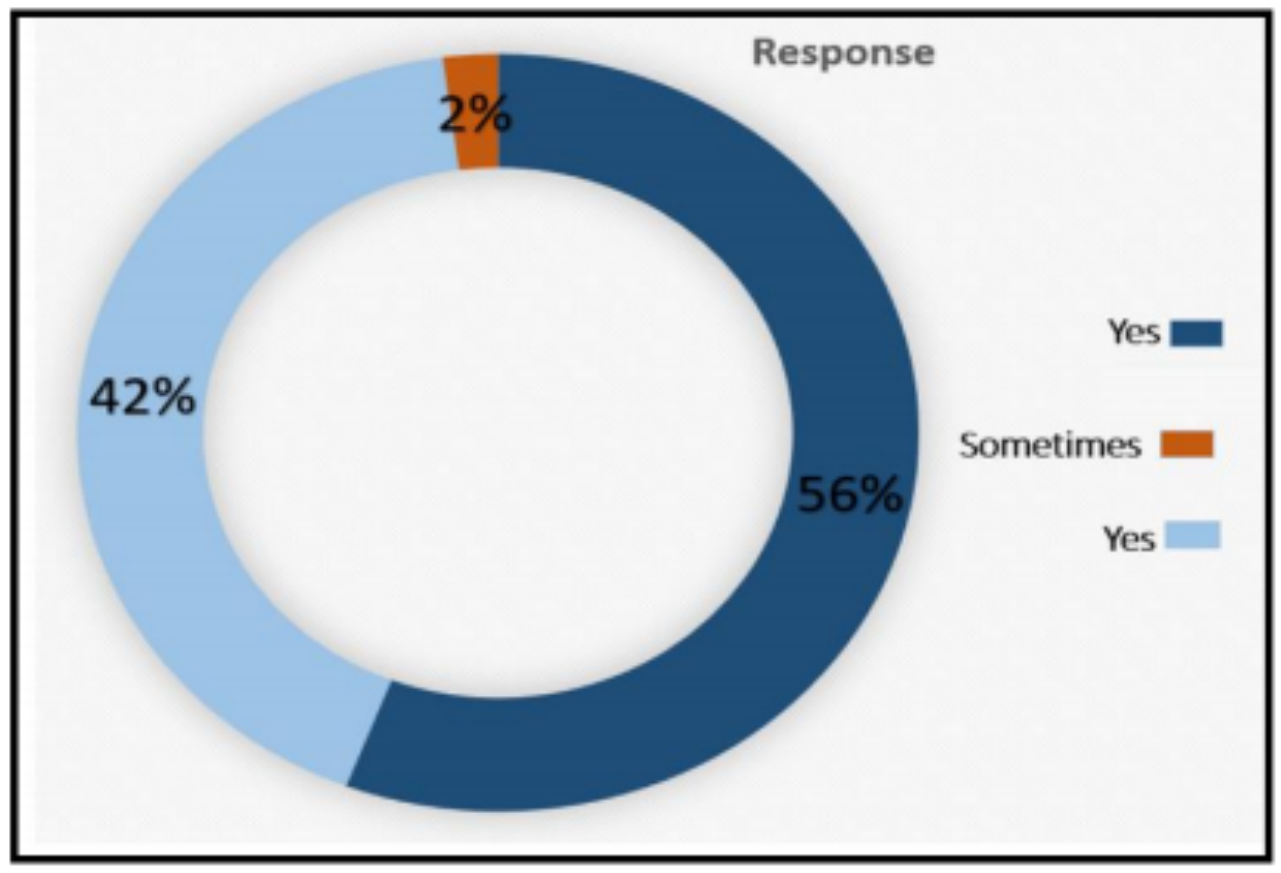

**Figure 3: Should learners be allowed to use mobile phones at school**

## 6. Conclusion and Future Research

The policy to ban mobile phones in township schools of the Western Cape province of South Africa is ineffective to stop learners from bringing their mobile devices to school despite the belief by school authorities that the policy works. Rather, this research shows that the students are more concerned with crime (theft and robbery of their devices) and lack of money to purchase internet data bundles. This security concern (fear of theft and robbery) is the main reason informing their decisions not to bring their phones, instead of the school ban.

While the study shows that the students more students are opposed to the school phone ban (56%), they tend to leave their phones behind primarily because of their concern about the harsh socio-economic realities of living in poor urban post-apartheid South Africa (crime and economic want). A significantly high number of learners (40%) are self-reporting to be using their phones for over 5 hours daily. This indicates a level of unhealthy overuse (which could indicate addiction) that needs to be addressed. The research also shows that leaners are using their phones mainly for socialising (communicating and chatting with friends and family via WhatsApp, Facebook and voice), browsing the internet for general (non-school related) information as well as using the phone as an entertainment and recreational device (music and taking pictures) and not primarily for educational purposes. It is therefore recommended that parents and school authorities should focus on addressing crime and safety primarily as that seems to be the student's primary concern. Furthermore, it is recommended that schools continuously make students aware about the dangers of mobile phone overuse.

A future research area is to understand student perceptions of mobile phone overuse and addiction as well helping students in economically disadvantaged areas to discover more educational uses of their mobile phones. This will help parents and learners to begin a conversation about the healthy use of mobile phone devices as well as encourage a more educationally beneficial use of technology by the learners.

# 7. References

Balakrishnan, V., & Raj, R. G. (2012). Exploring the relationship between urbanized Malaysian youth and their mobile phones: A quantitative approach. Telematics and Informatics, 29(3), 263–272.

Chigona, A., Chigona, W., Kayongo, P., & Kausa, M. (2010). An empirical survey on domestication of ICT in schools in disadvantaged communities in South Africa Agnes Chigona , Wallace Chigona , Patrick Kayongo and Moses Kausa. International Journal of Education and Development Using Information and Communication Technology, 6(2), 21–32.

Chigona, W., Kankwenda, G., & Manjoo, S. (2008). The uses and gratifications of mobile internet among the South African students. PICMET: Portland International Center for Management of Engineering and Technology, Proceedings, 10(September 2008), 2197– 2207.

Conradie, D., & Roodt, J. (2013). Realities versus ideals with redard to e-learning in South Africa, (2001), 2001–2003.

Corbeil, J. R., & Valdes-Corbeil, M. E. (2007). Are you ready for mobile learning? EDUCAUSE Quarterly, 30(2), 51–58.

Eybers, S., & Giannakopoulos, A. (2015). The Utilisation of Mobile Technologies in Higher Education: Lifebuoy or Constriction? In International Conference on Mobile and Contextual Learning (pp. 300-314). Springer International Publishing.

Ferriter, W. M., & Garry, A. (2010b). Teaching the iGeneration: 5 Easy Ways to Introduce Essential Skills With Web 2.0 Tools, 242.

Furze, T. (2014). E-Safety for the i-Generation: combatting the misuse and abuse of technology in schools. Educational Psychology in Practice, 30(3), 324–325.

Gao, Q., Yan, Z., Wei, C., Liang, Y., & Mo, L. (2017). Three different roles, five different aspects: Differences and similarities in viewing school mobile phone policies among teachers, parents, and students. Computers and Education, 106, 13–25.

Gao, Q., Yan, Z., Zhao, C., Pan, Y., & Mo, L. (2014). To ban or not to ban: Differences in mobile phone policies at elementary, middle, and high schools. Computers in Human Behaviour, 38, 25–32.

Hartnell-Young, E., & Vetere, F. (2008). A means of personalizing learning: incorporating old and new literacies in the curriculum with mobile phones. Curriculum Journal, 19(4), 283–292. Hong, F. Y.,

Chiu, S. I., & Huang, D. H. (2012). A model of the relationship between psychological characteristics, mobile phone addiction and use of mobile phones by Taiwanese university female students. Computers in Human Behavior, 28(6), 2152–2159.

Holfeld, B. (2012). Middle school students' perceptions of and responses to cyber bullying. Journal of Educational Computing Research, 46(4), 395–413.

Ives, E. a. (2012). iGeneration: The Social Cognitive Effects of Digital Technology on Teenagers. Online Submission, (October), 1–107.

Jahn, B., & Kunz, W. (2012). How to transform consumers into fans of your brand. Journal of Service Management, 23(3), 344–361.

Jones, C., Ramanau, R., Cross, S., & Healing, G. (2010). Net generation or Digital Natives: Is there a distinct new generation entering university? Computers and Education, 54(3), 722– 732.

Lederer, K. (2012). Pros and Cons of Social Media in the Classroom -- Campus Technology.

Lenhart, A., Ling, R., Campbell, S., & Purcell, K. (2010). Teens and Mobile Phones it as the centrepiece of their communication. Pew Internet & American Life Project, 1–114.

Maphalala, M. C., & Nzama, M. V. (2014). The Proliferation of Cell Phones in High Schools: The Implications for the Teaching and Learning Process. Mediterranean Journal of Social Sciences, 5(3), 461–466.

Murphy, R. (2015). in brief... Phone home: should mobiles be banned in schools? 10–11.

Nawaz, S., & Ahmad, Z. (2012). Statistical Study of Impact of Mobile on Student's Life, 2(1), 43– 49.

Ndlovu, N. S., & Lawrence, D. (2012). The quality of ICT use in South African classrooms. North, D.,

North, D., Johnston, K., & Ophoff, J. (2014). The use of mobile phones by South African university students. Issues in informing science and information technology, 11(2), 115-138.

O'Bannon, B. W., & Thomas, K. (2014). Teacher perceptions of using mobile phones in the classroom: Age matters! Computers and Education, 74, 15–25.

O'Bannon, B. W., & Thomas, K. M. (2015). Mobile phones in the classroom: Preservice teachers answer the call. Computers and Education, 85, 110–122.

O'Hagan, T. (2015). Mobility in Education: can mobile devices support teaching and learning in South Africa? 55–59.

Oblinger, D., Oblinger, J., L., & Lippincott, K. (2005). Educating the Net Generation. Educational Leadership (Vol. 56).

Philip, T., & Garcia, A. (2013). The importance of still teaching the iGeneration: New technologies and the centrality of pedagogy. Harvard Educational Review, 83(2), 300–320.

Porter, G., Hampshire, K., Milner, J., Munthali, A., Robson, E., de Lannoy, A., & Abane, A. (2016). Mobile phones and education in sub-Saharan Africa: from youth practice to public policy. Journal of international development., 28(1), 22-39.

Rambe, P., & Bere, A. (2013). Using mobile instant messaging to leverage learner participation and transform pedagogy at a South African University of Technology. British Journal of Educational Technology, 44(4), 544–561.

Ruggiero, T. E. (2000). Uses and gratifications theory in the 21st century. Mass Communication & Society, 3(1), 3–37.

Sanz-blas, R. C. C. R. S., Curras-perez, R., Ruiz-mafe, C., & Sanz-blas, S. (2014). approach from the uses and gratifications perspective Determinants of user behaviour and recommendation in social networks An integrative approach from the uses and gratifications perspective, 114(9), 1477–1498.

Sundari, T. T. (2015). Effects of mobile phone use on academic performance of college going young adults in India, 1(9), 898–905.

Takalani, T. (2008). Barriers To E-Learning Amongst Postgraduate Black Students in Higher, (February).

Traxler, J. (2009). Current state of mobile learning. In M. Ally (Ed.), Mobile learning: Transforming the delivery of education and training (pp. 247–264). Edmonton, Alberta, Canada: Athabasca Press.

Van Der Ross, D., and Tsibolane, P. (2017). The influence of teacher attitudes and beliefs on information and communications technology integration behavior in south African high schools. CONF-IRM 2017 Proceeding, p. 32.

Walls, E., Santer, M., Wills, G., & Vass, J. (2014). The dream plan: A Blupont stratergy for e-education provision in South Africa. Igarss 2014, (1), 1–5.